\pdfoutput=1
\documentclass[%
 reprint,
 amsmath,amssymb,
 aps,
]{revtex4-2}

\usepackage{graphicx}
\usepackage{dcolumn}
\usepackage{bm}
\usepackage{amsmath}
\usepackage{float}
\usepackage[T1]{fontenc}
\usepackage[utf8]{inputenc}
\usepackage{hyperref}
\hypersetup{
    colorlinks,
    citecolor=blue,
    linkcolor=blue,
    urlcolor=blue
}

\newcommand{\dd}{\mathrm{d}}

\begin{document}

\preprint{APS/123-QED}

\title[Photon rings of spherically symmetric non-Kerr black holes]{Photon rings of spherically symmetric black holes and robust tests of non-Kerr metrics}

\author{Maciek Wielgus}
 \email{maciek.wielgus@gmail.com}
	\affiliation{Black Hole Initiative at Harvard University, 20 Garden Street, Cambridge, MA 02138, USA}
		\affiliation{Center for Astrophysics  $|$ Harvard \& Smithsonian, 60 Garden Street, Cambridge, MA 02138, USA}

\date{\today}

\begin{abstract}
Under very general assumptions on the accretion flow geometry, images of a black hole illuminated by electromagnetic radiation display a sequence of \textit{photon rings} (demagnified and rotated copies of the direct image) which asymptotically approach a purely theoretical \textit{critical curve} -- the outline of the black hole photon shell.
To a distant observer, these images appear dominated by the direct emission, which forms a ring whose diameter is primarily determined by the effective radius of the emitting region.
For that reason, connecting the image diameter seen by a distant observer to the properties of the underlying spacetime crucially relies on a calibration that necessarily depends on the assumed astrophysical source model.
On the other hand, the diameter of the photon rings depends more on the detailed geometry of the spacetime than on the source structure.
As such, a photon ring detection would allow for the spacetime metric to be probed in a less model-dependent way, enabling more robust tests of General Relativity (GR) and the Kerr hypothesis.
Here we present the photon ring structure of several spherically symmetric black hole spacetimes and perform comparisons with the Schwarzschild/Kerr case.
We offer our perspective on future tests of the spacetime metric with photon rings, discussing  challenges and opportunities involved.
\end{abstract}

\maketitle

\section{Introduction}
\label{sec:intro}

\begin{figure*}[t!]
    \centering
     \includegraphics[width=0.99\linewidth]{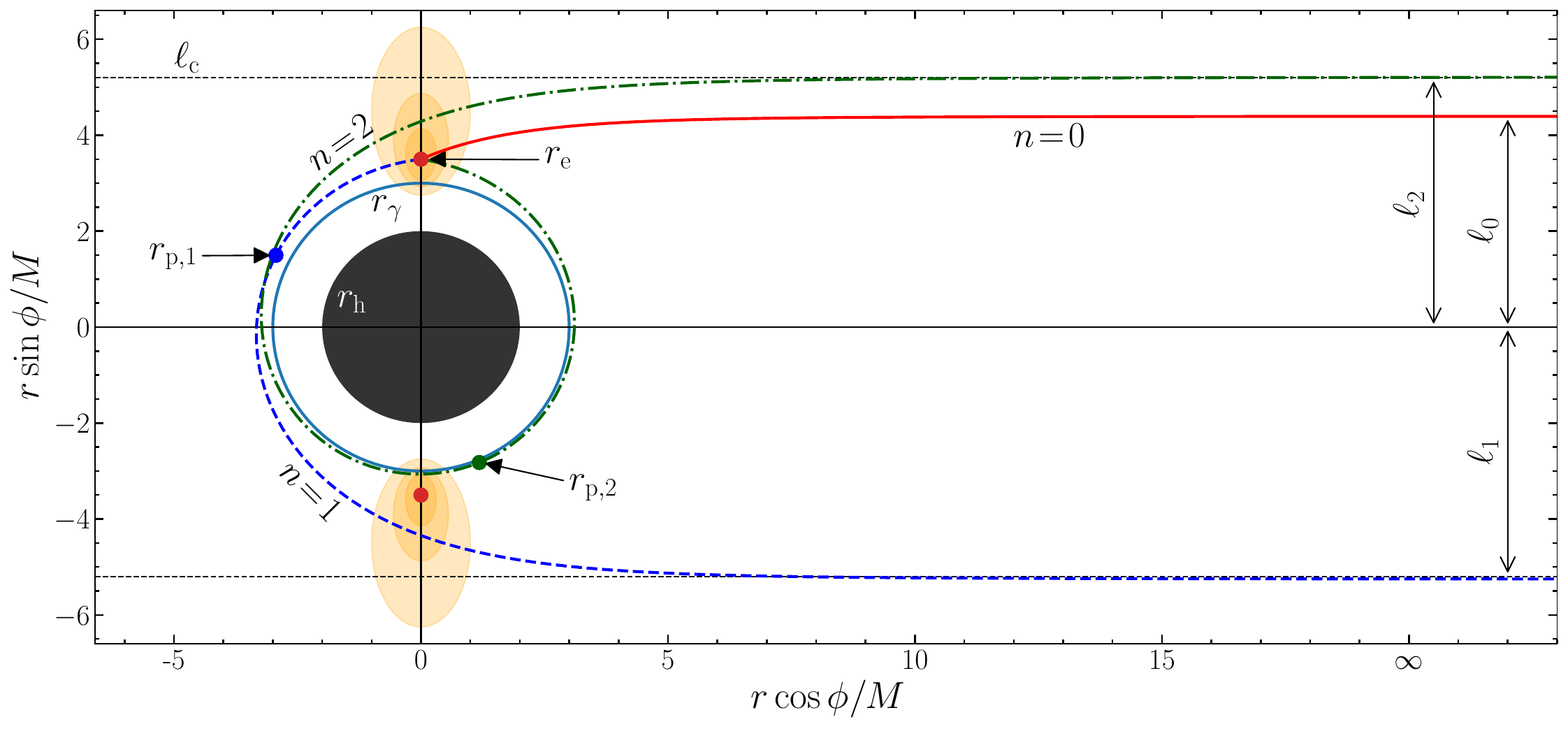}
    \caption{A simple model of the emission around a spherically symmetric static black hole. The observer is located to the right, at $r\rightarrow \infty$, viewing the accretion disk face-on. Black hole has a horizon radius $r_{\rm h}$, there is a photon sphere located at $r_\gamma$ (blue circle). The emission from the accretion disk (represented with orange ellipses) is dominated by the flux emitted at the effective radius $r_{\rm e}$. The photons emitted at $r_{\rm e}$ may reach the observer along infinite number of trajectories, corresponding to adding half loops around the photon sphere. First 3 such trajectories are shown. They correspond to the direct image ($n=0$), and first and second photon ring ($n=1$ and $n=2$). Radii of $n$-th photon rings in the observer's plane converge rapidly to the critical impact parameter, $\ell_n \rightarrow \ell_{\rm c}$, which defines the critical curve on the observer's screen.
    }
    \label{fig:introfigure}
\end{figure*}

The observational appearance of a black hole (BH) has been studied extensively for many years from a purely theoretical standpoint \citep{Darwin1959,Bardeen1973,Luminet1979,Falcke2000}. This appearance depends on numerous astrophysical details, such as geometry and optical depth of the emission region \citep{Gralla2019,Narayan2019,Vincent2020}. The feature depending exclusively on the spacetime geometry is the critical curve \citep{Gralla_lensing}, an outline of the photon shell on a distant observer's screen. However, since the null geodesics comprising the photon shell are unstable orbits \citep{Teo2003}, the critical curve remains a purely mathematical concept. Unless the accretion flow itself is spherically symmetric \citep{Falcke2000,Narayan2019} (which is typically not an astrophysically plausible scenario), the image of a BH contains a sequence of discrete photon rings \citep{Gralla_lensing,Johnson2020}.
Each photon ring is a copy of the direct image of the emission region, corresponding to photon trajectories lensed around the photon shell, with the $n$-th photon ring photons executing $n/2$ loops before reaching the observer's screen \citep{Gralla2019,Johnson2020}, see Fig. \ref{fig:introfigure}. This sequence of photon rings quickly converges to the critical curve, and the diameter and shape of the $n\!=\!2$ photon ring is already strongly dominated by the spacetime geometry -- hence it can be used for precise tests of the theory of gravity, with only very general assumptions on the astrophysical model of the accretion flow \citep{Johnson2020,Gralla_n2}. In the case of non-Kerr spacetimes, most of the previous research has focused on the theoretical critical curve (commonly referred to as \textit{shadow}, occasionally also as \textit{photon ring}, which has a different meaning in this paper), which has been characterized for a broad range of non-Kerr spacetimes, e.g., \citep{Zakharov1994,Bambi2013,Younsi2016,Medeiros2020,Wielgus2020worm,Bacchini2021}. Relatively less work has been dedicated to the observable photon ring structure. Additionally, the observable ring-like direct emission feature is commonly conflated with the critical curve in the literature. One of the aims of this paper is to clarify that issue.

As noticed by \citep{Johnson2020}, the technique of very long baseline interferometry (VLBI) is uniquely suited to probe the photon ring structure of black hole images. As sharp image domain features, photon rings die-off slowly in the Fourier domain. Hence, they tend to dominate the interferometric signal in the range of very high spatial frequencies, corresponding to baseline
lengths of tens and hundreds billion wavelengths. 
While such long baselines were never observed so far and this feat will surely require significant progress and new developments in the field of space VLBI, the projects that would enable robust detection of the photon ring structure are already being discussed and proposed, e.g., \citep{Gralla_n2,Johnson2020,Haworth2019,Pesce2019,Vincent2020, Gurvits2021}, and algorithms to enable photon ring detection in VLBI data are already being implemented and tested with synthetic data \citep{Gralla_n2,Broderick2020,Broderick2021}.

Only recently first resolved observational images of a BH were obtained by the Event Horizon Telescope VLBI array (EHT; \citep{EHT2019p1,EHT2019p2,EHT2019p3,EHT2019p4,EHT2019p5,EHT2019p6}), presenting the supermassive BH M\,87$^*$ in the center of the M\,87 galaxy with a resolution corresponding to $\sim 3 r_{\rm S}$ \citep{EHT2019p3,EHT2019p4,EHT2019p5}, where $r_{\rm S}$ is the Schwarzschild radius.
The measured diameter of a bright ring feature found in the EHT images was used to test the null hypothesis of consistency with the Kerr metric \citep{Psaltis2019,EHT2019p6} under the BH mass prior derived from stellar dynamics \citep{Gebhardt2011}. To formulate this test, the scaling relationship between the ring diameter and the BH mass assuming Kerr metric was evaluated using 103 synthetic images generated with general relativistic magnetohydrodynamic simulations (GRMHD; \citep{EHT2019p5,EHT2019p6}). These synthetic images, representing variety of viable theoretical GRMHD models of BH accretion, were used to perform diameter-mass calibration and quantify the null hypothesis test. The results of this test were later on 
reinterpreted as EHT constraints on the diameter of the critical curve to be 17\,\% consistent with that of the Kerr spacetime \citep{Dimitrios2020,Kocherlakota2021}. 
However, the exact value of the uncertainty depends on the selection of simulated images and underlying parameters of the models used to perform calibration, and since the images are computed under Kerr spacetime assumption, the uncertainties are not directly transferable to tests of non-Kerr spacetimes. Thus, the 17\,\% uncertainties assumed by \citep{Dimitrios2020,Kocherlakota2021} can only be considered a proxy to demonstrate the principle of testing spacetime geometry with the EHT observations, but one should be extremely careful with the quantitative interpretation of any constraints provided in this way. 

While numerical GRMHD constitutes the most sophisticated modeling framework we have at our disposal to interpret the EHT observations, there are good reasons to question whether GRMHD simulations can represent the appearance of an accreting BH accurately enough to be used for a reliable diameter-mass calibration (the fact that certain GRMHD models \textit{are consistent} with the EHT observations only means fulfilling one necessary condition). Unlike objections related to the application of the diameter-mass calibration procedure to non-Kerr spacetimes, these concerns challenge the conclusions obtained within Kerr paradigm as well. Some of these concerns were enumerated by \citep{Gralla_canEHT} and we will not repeat them here. Instead, we focus on demonstrating how the measurement of the direct emission-dominated ring image diameter translates (under a prior on BH mass) to a measurement of the effective emission region size, largely independent of the underlying metric, while photon rings exhibit much more sensitivity to the underlying spacetime geometry. To that end we investigate the photon ring structure of several spherically symmetric BH spacetimes (spacetimes with an event horizon) and compare them with Schwarzschild/Kerr metrics.

\section{Astrophysical constraints}
\label{sec:astrophysical_constraints}

VLBI observations have a potential to resolve supermassive BHs in the centers of galaxies. Two targets of largest angular diameter are M\,87$^*$ \citep{EHT2019p1} and Sagittarius A$^*$ (Sgr A$^*$), the latter being the supermassive black hole in the center of our own galaxy \citep{Doeleman2008}. A number of other supermassive BH sources could be resolved in a foreseeable future with space VLBI \citep{Haworth2019}. Apart from the concerns about the angular resolution, a potential source must emit sufficiently strong radiation at relevant frequencies ($10^{11} - 10^{12}$\,Hz). Spectral observations unambiguously indicate \citep[e.g.,][]{Narayan1995,DiMatteo2003} that the accretion flow in the objects of interest corresponds to a hot and radiatively inefficient accretion flow (RIAF; also commonly referred to as the advection dominated accretion flow - ADAF) model \citep{Rees1982,Narayan1994,Abramowicz1995,Yuan2014,Bronzwaer2021}, where the mass accretion rate is very low, and the radiation observed by VLBI instruments originates from a synchrotron process. A classic thin accretion disk model \citep{Novikov1973} represents a very different accretion regime (geometrically thinner, optically thicker, less hot, more dense, emitting thermal radiation) and therefore \emph{should not} be used for quantitative interpretations of the VLBI images of supermassive BHs. A particularly relevant difference seems to be that in the analytic thin disk model radiative flux terminates at the innermost stable circular orbit (ISCO), which is not the case for the RIAF models -- thus the VLBI images of BHs are not expected to necessarily retain clear information about the ISCO location. While we expect the RIAF synchrotron emission at all radii above the event horizon (located at radius $r_{\rm h}$), in the horizon limit photons are infinitely redshifted. Hence, for the axisymmetric flow, regardless of the details of the accretion model, there must be an effective radius of emission $r_{\rm e} > r_{\rm h}$, where majority of photons reaching the distant observer originate from. 

In the simple model discussed in this paper we technically assume that all of the emission originates at particular $r_{\rm e}$, only computing images of an infinitely thin ring of radius $r_{\rm e}$ and disregarding the spatially extended character of the emission in a more physically realistic scenario. A definition of $r_{\rm e}$ may be more ambiguous in a realistic case, and depending on the details of the emission profile. However, that is exactly the reason why observing photon rings, far less dependent on $r_{\rm e}$ than the direct image, is a tantalizing perspective.

A separate concern is related to the optical depth -- the observations need to penetrate the outer layers of the accretion flow and allow to peer into its innermost region in the immediate vicinity of the event horizon. Synchrotron emission has two regimes, self-absorbed optically thick regime at lower observing frequencies, and optically thin regime for higher frequencies with a steep decrease of flux as a function of frequency \citep{Rybicki}. These regimes are separated by the emission maximum at a turn-over frequency $\nu_t \propto M_{\rm BH}^{-1} \dot{M}^{2/3}$ \citep{Falcke2004} for black hole mass $M_{\rm BH}$ and mass accretion rate $\dot{M}$. Hence, the observing frequency needs to be close to the turn-over frequency $\nu_t$ to reach the optically thin regime, but still collect sufficient amount of flux to enable VLBI detections. Fortunately, that seems to be the case for both M\,87$^*$ and Sgr\,A$^*$ at the current EHT observing frequency of $2.3\times10^{11}$\,Hz \citep{Doeleman2008, EHT2019p1}.

\begin{figure}[b!]
    \centering

     \includegraphics[width=0.99\linewidth]{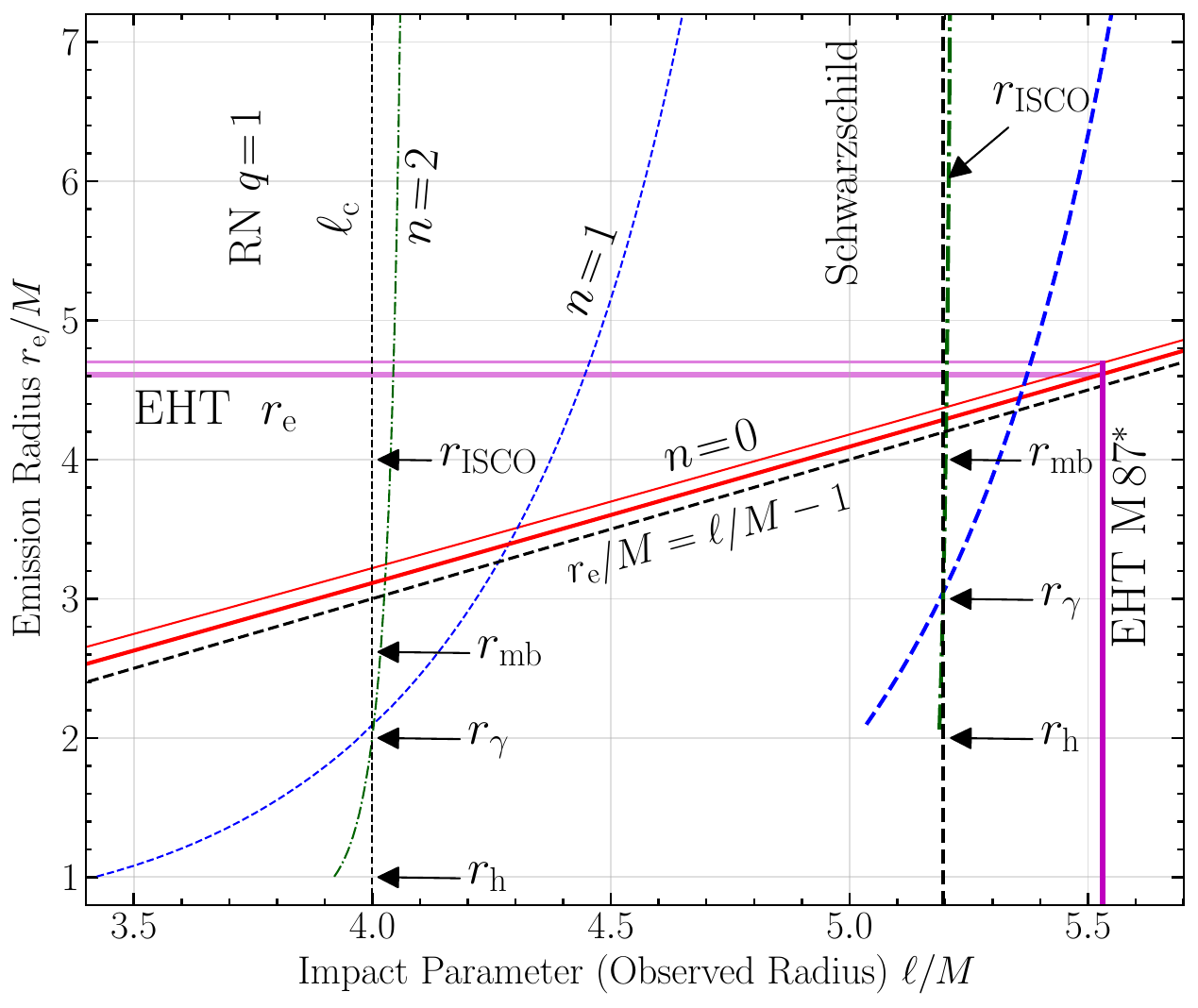}
    \caption{Transfer functions comparison between Schwarzschild (thick lines) and maximally charged Reissner-Nordstr\"om (thin lines) spacetimes. Assuming known $M/D$, EHT observations of M\,87$^*$ interpreted as a ring dominated by the $n\!=\!0$ emission component allow to calculate the effective emission radius $r_{\rm e}$ (horizontal magenta line). The EHT ring diamater can be explained by $r_{\rm e} = 4.61\,M$ in Schwarzschild or by $r_{\rm e} = 4.70\,M$ in RN spacetime. Unlike direct emission $n=0$ transfer function, higher order transfer functions are separated along the observed radius horizontal axis, and hence may be used as a robust discriminant between the two spacetimes. }
    \label{fig:transferfunctions}
\end{figure}

The fundamental physics behind RIAF flow and synchrotron radiation does not depend on the spacetime metric. However, the same can not be said about the detailed dynamics of the accretion flow and relative importance of the gravitational redshift, both influencing the location of the effective flux maximum $r_{\rm e}$. The location of $r_{\rm e}$ is also strongly affected by physics of plasma dissipation and electron heating, which remains highly uncertain in the relevant regime. While these aspects of physics are not directly dependent on the spacetime geometry, they remain insufficiently understood and are prescribed rather than self-consistently calculated in the GRMHD simulations \citep{EHT2019p5}. Hence, the impact of the accretion model on the $r_{\rm e}$ may easily be mistaken for the effect of the spacetime geometry. This is illustrated in an example shown in Fig. \ref{fig:transferfunctions}. We consider the case of a spherically symmetric spacetime. Hence, critical curve in the observer's screen corresponds to a circle with radius equal to a critical impact parameter $\ell_{\rm c}$ (the exact scaling requires the knowledge of the mass and distance of the observed BH), see Fig. \ref{fig:introfigure}. We also limit ourselves to the case of an observer viewing the accretion disk around a static black hole face-on (along the spin axis of the accretion disk). In that idealized case photon rings are simply concentric circles in the distant observer's screen. Figure \ref{fig:transferfunctions} presents transfer functions \citep{Gralla2019,Gralla_lensing} connecting the emission radius $r_{\rm e}$ (in BH mass units) and the impact parameter $\ell_n$ of $n$-th photon ring for Schwarzschild and Reissner-Nordstr\"om (RN) metric with charge parameter $q\!=\!1$ (see Section \ref{sec:spacetimes}). \citep{Kocherlakota2021} excluded RN metric with $q\!=\!1$ based on the aforementioned EHT $17\,\%$ Kerr-GRMHD-tuned error budget. Let us critically examine this conclusion. For simplicity, let us assume that the mass and distance of M\,87$^*$
are known perfectly well, with no uncertainties, yielding ratio of $M/D = 3.8\,\mu$as (\citep{EHT2019p1}; accounting for the additional uncertainties would only strengthen our argument). Then the EHT observation of a 42\,$\mu$as ring gives the impact parameter of $\ell_{\rm EHT}= 5.53 M$, a thick vertical magenta line in Fig. \ref{fig:transferfunctions}. Since the observed emission is strongly dominated by the $n\!=\!0$ direct component \citep{Gralla2019, Johnson2020}, we can use the $n\!=\!0$ transfer function to translate this measurement to the effective radius of emission $r_{\rm e}$ equal to 4.61\,$M$ for Schwarzschild (thick horizontal magenta line), and 4.70\,$M$ for RN spacetime (thin horizontal magenta line, the details of the calculations are presented in Section \ref{sec:calculations}). These values are very close to one another, because transfer functions of a direct image ($n=0$, red tilted lines) are very similar in both spacetimes. In both cases they are well approximated by a simple linear relation (tilted dashed black line; \citep{Gralla_lensing,Gates2020})
\begin{equation}
r_{\rm e} \approx \ell - M \ .
\label{eq:adding_one}
\end{equation}
Hence, the argument against RN spacetime with $q=1$ hinges on the assertion that we understand the astrophysics of the accreting system well enough to confidently claim that $r_e$ in RN spacetime \textit{must} be of value lesser than $4.70 M$.

The position we take in this paper is that such claim is unsubstantiated, given the serious uncertainties about the physics governing the radiative flux profile (where is the effective maximum $r_{\rm e}$ is located) and lack of dedicated accretion simulations performed in the RN spacetime (or in non-Kerr BH spacetimes in general, with a notable exception of \citep{Mizuno2018}). On the other hand, notice that a detection of the $n\!=\!1$ (or, better, $n>1$) photon ring would immediately allow to distinguish between these two spacetimes (at least in the case of a known $M/D$), as there is no overlap between the observable impact parameters of $n\!=\!1$ photon rings in the two spacetimes (at least for $r_{\rm e}<20M$), \textit{irrespectively of the underlying astrophysics}.

\section{Non-Kerr models of black holes}
\label{sec:spacetimes}

To demonstrate the properties of photon rings and the direct emission ring, let us consider a general spherically symmetric spacetime with a metric in a form
\begin{equation}
    \dd s^2 = g_{\mu \nu} \dd x^{\mu} \dd x^{\nu} = -f(r) \dd t^2 + \frac{g(r)}{f(r)} \dd r^2 + r^2 \dd \Omega^2
\label{eq:metric}
\end{equation}
The square root of $f(r)$ is commonly referred to as a \textit{lapse function} and $\dd \Omega^2 = \dd \theta^2 + \sin^2 \theta \dd \phi^2$ corresponds to a line element on a unit sphere. Given the spherical symmetry we will restrict ourselves to $\theta= \pi/2$ plane, where $\dd \Omega = \dd \phi$. All solutions are parametrized with the ADM (Arnowitt-Deser-Misner) mass of the central object $M$. Our reference solution is that of a Schwarzschild spacetime, a special case of Kerr spacetime in the static limit (BH spin $a_* = a/M = 0$), with
\begin{equation}
   f_{\rm S}(r) = 1 - \frac{2M}{r} \ \ ; \ \  g_{\rm S}(r) \equiv 1 \ .
\end{equation}
Following the Birkhoff theorem, it constitutes a unique spherically symmetric static vacuum solution to Einstein field equations (EFE) of GR, with a horizon at the Schwarzschild radius $r_{\rm h} = r_{\rm S}= 2M$. 
In this paper we discuss the photon ring structure of several other spacetimes, with metrics parametrized with a single charge parameter $q$, cast in the units of $M$, reducing to a Schwarzshild solution for $q=0$. Our selection of models is by no means complete and roughly corresponds to a subset of spacetimes considered in \citep{Kocherlakota2021,Kocherlakota2020}. In this section we briefly introduce these spacetimes.

\begin{figure*}[t!]
    \centering

     \includegraphics[width=0.325\linewidth]{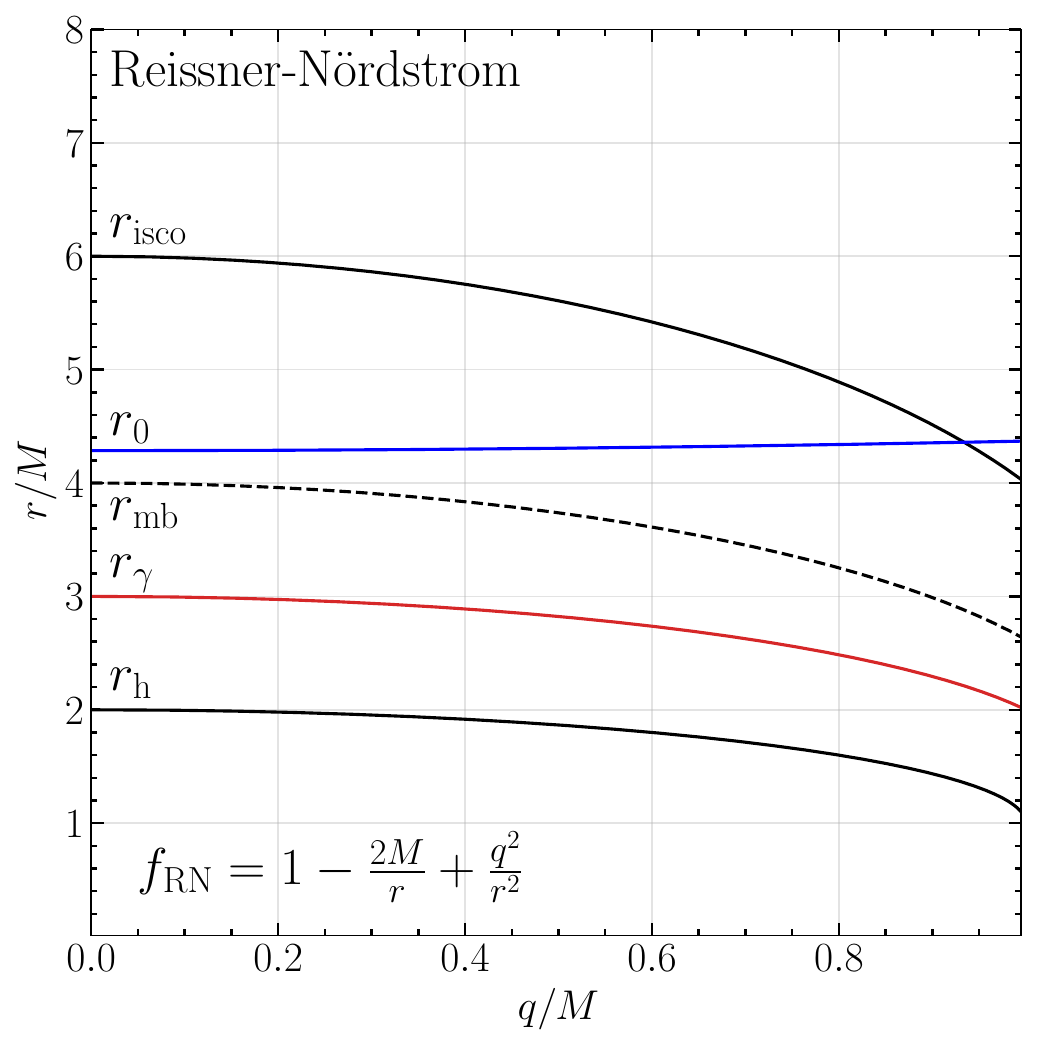}
     \includegraphics[width=0.325\linewidth]{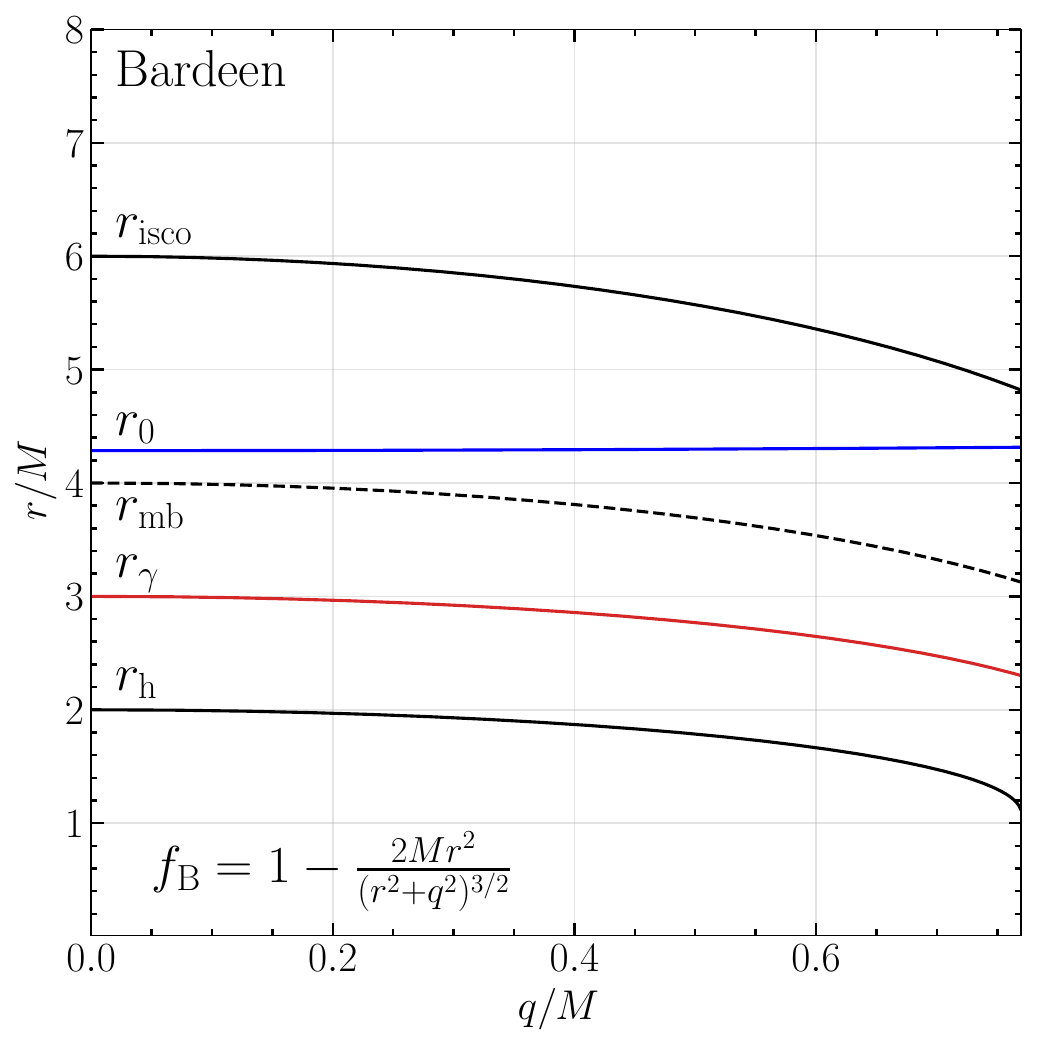}
     \includegraphics[width=0.325\linewidth]{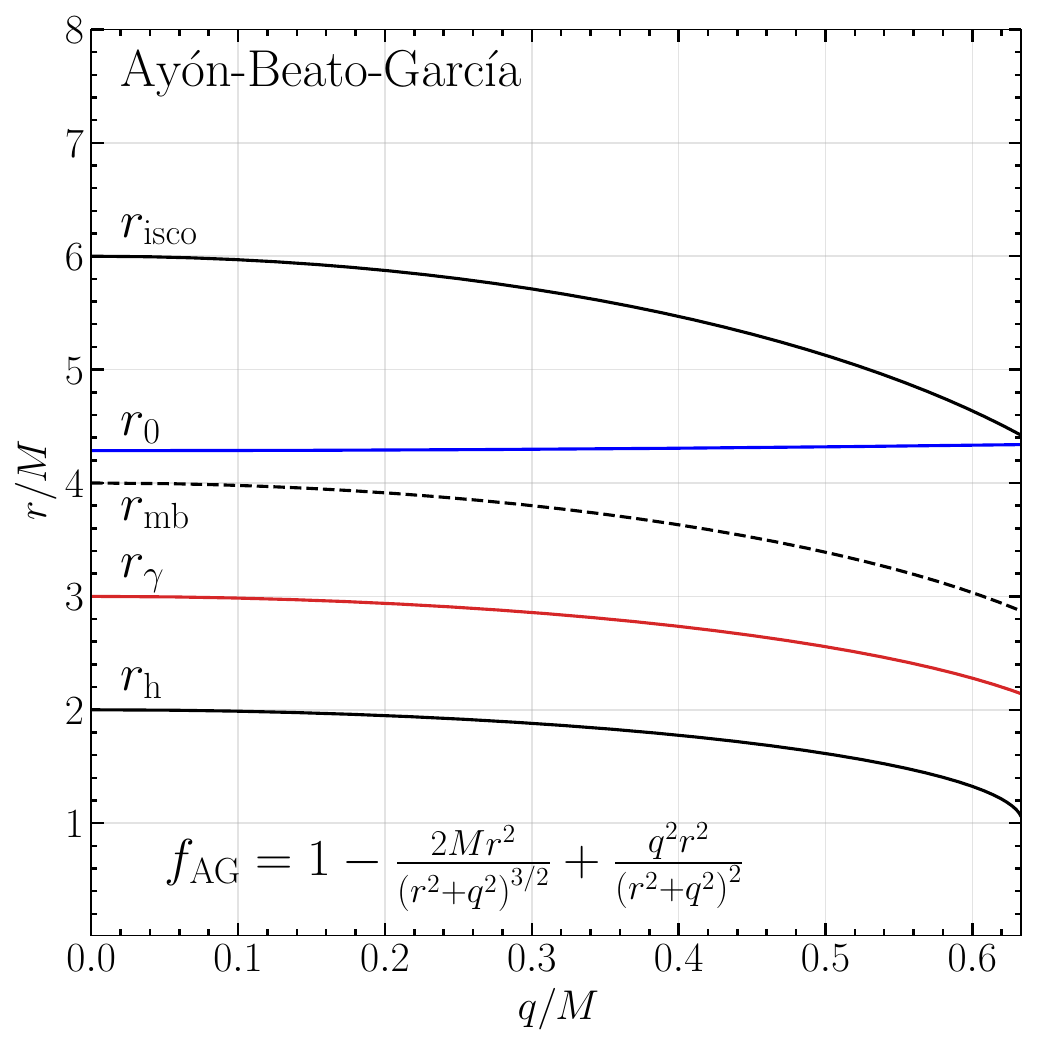}
     \includegraphics[width=0.325\linewidth]{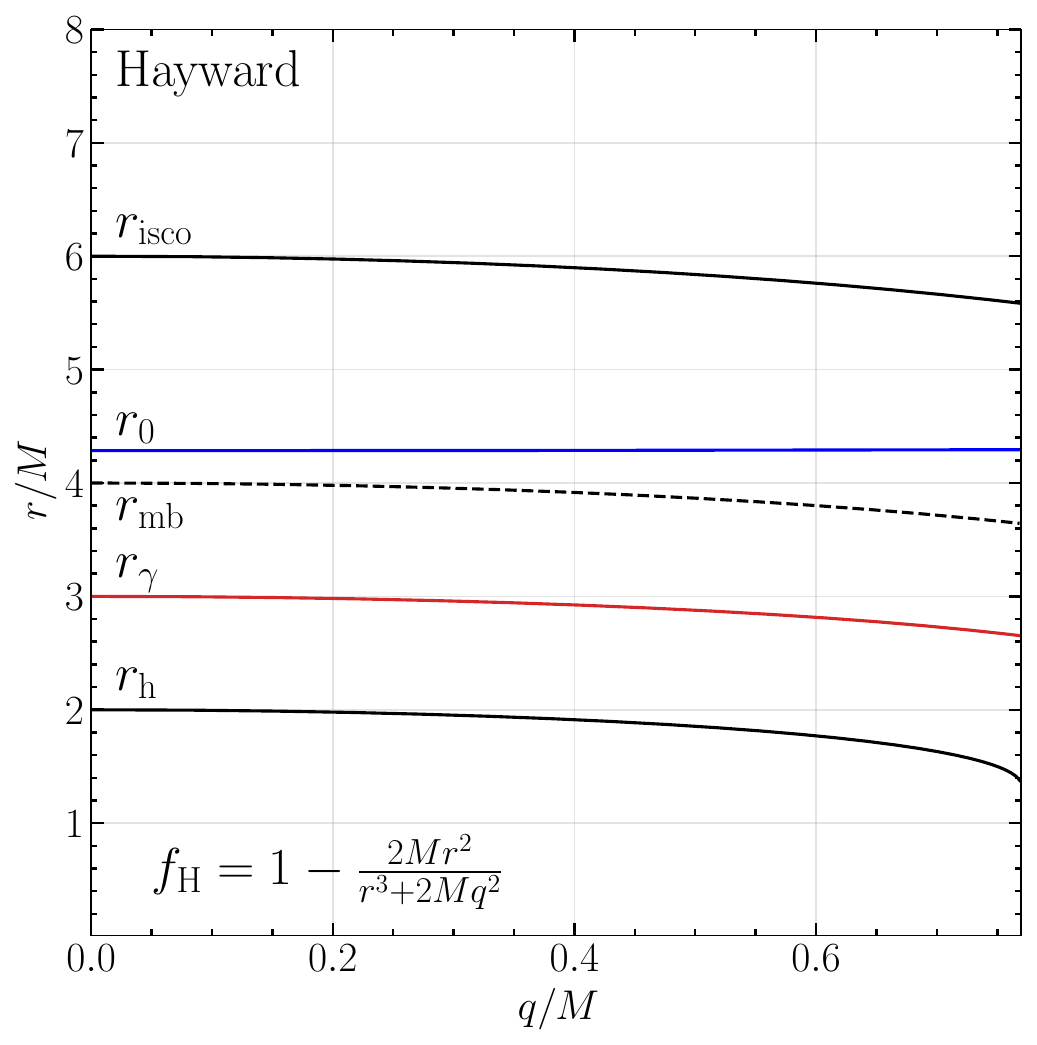}
     \includegraphics[width=0.325\linewidth]{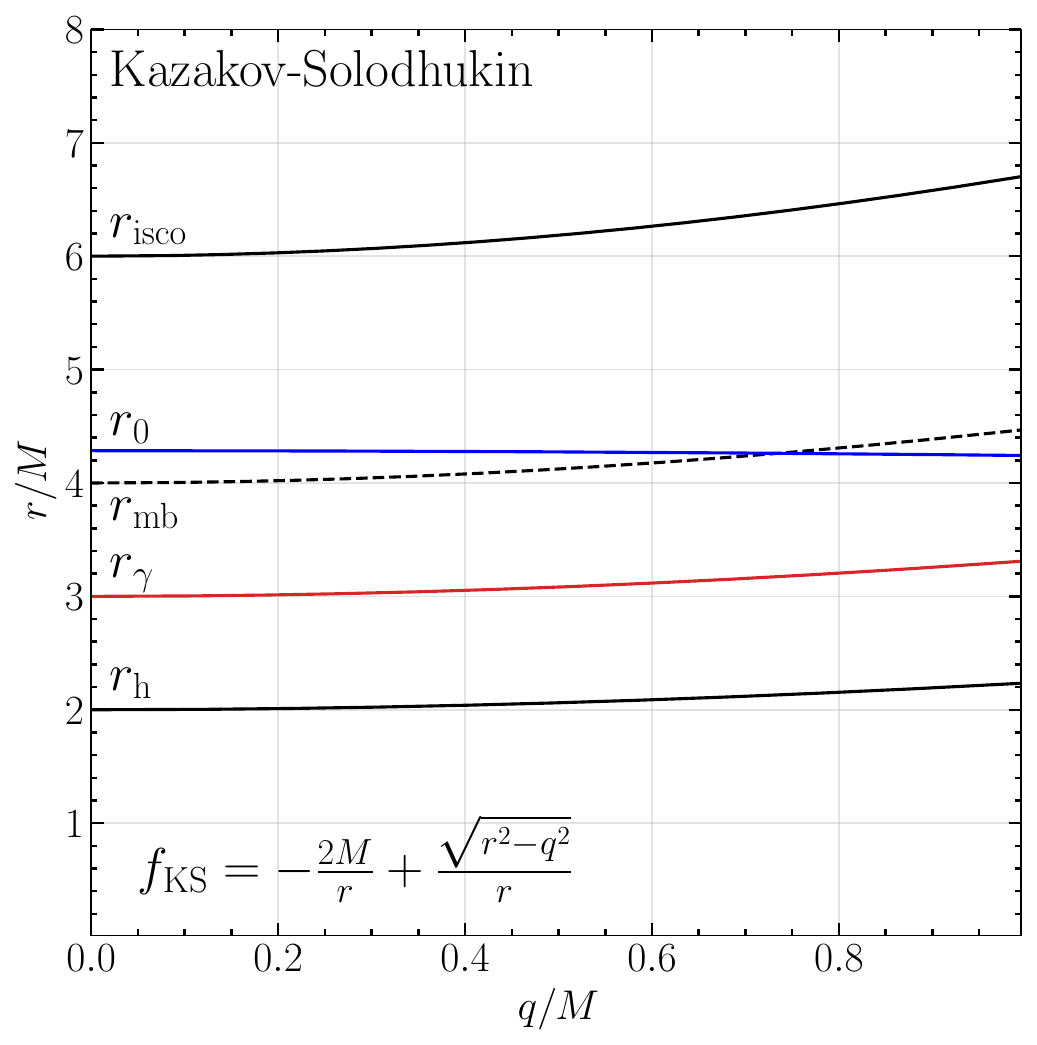}
     \includegraphics[width=0.325\linewidth]{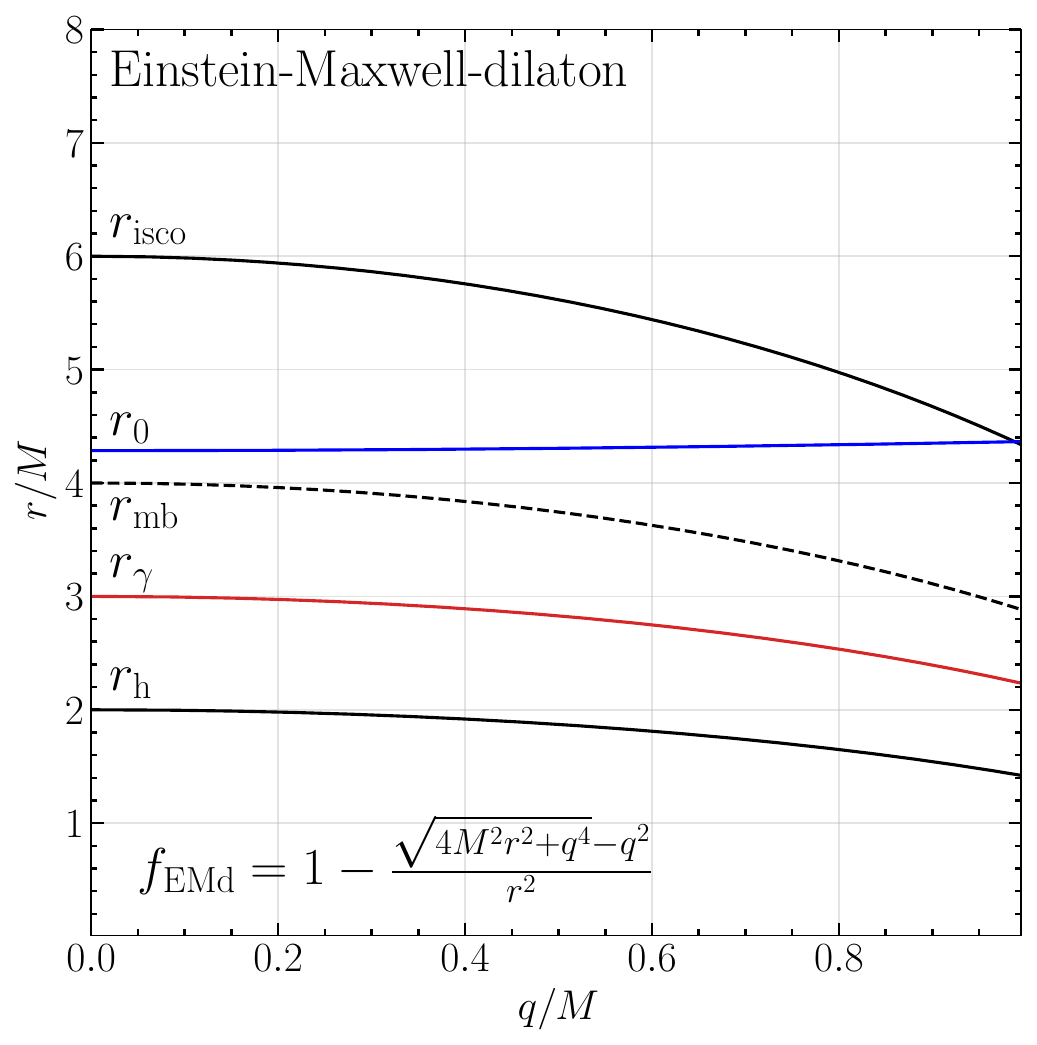}
    \caption{Geodesic structure of several spherically-symmetric BH spacetimes. The radius $r_0$ corresponds to the emission radius $r_{\rm e}$ for which the direct emission ring image has a radius corresponding to that of a Schwarzschild critical curve ($3\sqrt{3}M$). Note that $r_0$ is only very weakly dependent on the spacetime metric and is consistent with $4.3 \pm 0.1 M$ for all considered cases.
   }
    \label{fig:structure_BH}
\end{figure*}

\textit{Reissner-Nordstr\"om metric} (RN) is a classic solution to electrovacuum EFE, representing a static electrically charged black hole in GR. While there exist astrophysical arguments against accumulation of large electric charge $q$ by BHs, the RN solution remains a useful framework to test the $\propto r^{-2}$ deviations in $f(r)$. Moreover, RN solution also arises in the Randall-Sundrum model of the brane cosmology \citep{Dadhich2000}, where $q$ is reinterpreted as a tidal charge, representing gravitational effects from the higher dimensions. 
RN solution is defined with
\begin{equation}
   f_{\rm RN}(r) =  1 - \frac{2M}{r} + \frac{q^2}{r^2} \ \ ; \  \ g_{\rm RN}(r) \equiv 1 \ .
\end{equation}
Diameter of the RN black hole critical curve 
was studied by \citep{Zakharov2014}.

\textit{Regular black hole} solutions attempt to avoid the presence of the central singularity, where the curvature is unbounded and GR loses its validity. First such solution corresponded to a magnetically charged BH and was proposed by Bardeen \citep{bardeen1968} with
\begin{equation}
    f_{\rm B}(r) = 1 - \frac{2Mr^2}{\left(r^2+ q^2\right)^{3/2}} \ \ ; \ \ g_{\rm B}(r) \equiv 1 \ .
\end{equation}
This postulated metric inspired regular BH solutions of EFE coupled to nonlinear electrodynamics, e.g., by Ay\'{o}n-Beato and Garc\'{i}a \citep{Beato1998},
\begin{equation}
    f_{\rm AG}(r) = 1 - \frac{2Mr^2}{\left(r^2+ q^2\right)^{3/2}}  + \frac{q^2r^2}{\left(r^2+ q^2\right)^{2}} \ \ ; \ \ g_{\rm AG}(r) \equiv 1,
\end{equation}
and by  Hayward \citep{Hayward2006}, with
\begin{equation}
    f_{\rm H}(r) = 1 - \frac{2Mr^2}{r^3+ 2q^2M}  \ \ ; \ \ g_{\rm H}(r) \equiv 1 \ .
\end{equation}
The critical curves of regular black holes were studied by \citep{Abdujabbarov2016} and \citep{Stuchlik2019}.

\textit{Kazakov-Solodukhin black hole} solution (KS) considers the deformation of the Schwarzschild solution due to spherically symmetric quantum fluctuations, yielding a metric with
\begin{equation}
    f_{\rm KS}(r) =  - \frac{2M}{r} + \frac{1}{r} \left(r^2 - q^2 \right)^{1/2}  \ \ ; \ \ g_{\rm KS}(r) \equiv 1 .
\end{equation}
The critical curve diameter of the KS black hole was considered by \citep{Konoplya2020}.

\textit{Einstein-Maxwell-dilaton black hole} (EMd) is an electromagnetically charged BH solution in the Einstein-Maxwell-dilaton theory, where the dilaton field couples to the electromagnetic field  \citep{Gibbons1988,Garfinkle1991}. Following \citep{Kocherlakota2020} we consider a following static EMd metric
\begin{align}
    &f_{\rm EMd}(r) = 1 - \frac{\sqrt{4 M^2r^2 + q^4} -q^2}{r^2} \ ; \nonumber \\ &g_{\rm EMd}(r)  = \frac{4 M^2 r^2}{4 M^2 r^2 + q^4} .
\end{align}
The critical curves of EMd BHs has been studied in \citep{Wei2013}.

Differences between these spacetimes are shown in Fig. \ref{fig:structure_BH}, where we compare their properties by plotting characteristic radii relevant for the geodesic dynamics of particles \citep{Bardeen1972} -- event horizon radius $r_{\rm h}$, photon sphere radius $r_{\gamma}$, marginally bound orbit radius $r_{\rm mb}$, and ISCO radius $r_{\rm isco}$ as a function of the charge parameter $q$. The relevant radii can be found as roots of the following equations
\begin{align}
    & r_{\rm h} \, : \, f = 0 , \\
    & r_{\rm \gamma} \, : \, f'r - 2 f = 0 , \label{eq:rgamma} \\
    & r_{\rm mb} \, : \, f'r - 2 f (1- f)= 0 ,  \\
        & r_{\rm isco} \, : \, f''r - 2 r (f')^2 + 3 f f'\!=\!0  \ .
\end{align}

\begin{figure}[t!]
    \centering

     \includegraphics[width=0.9\linewidth]{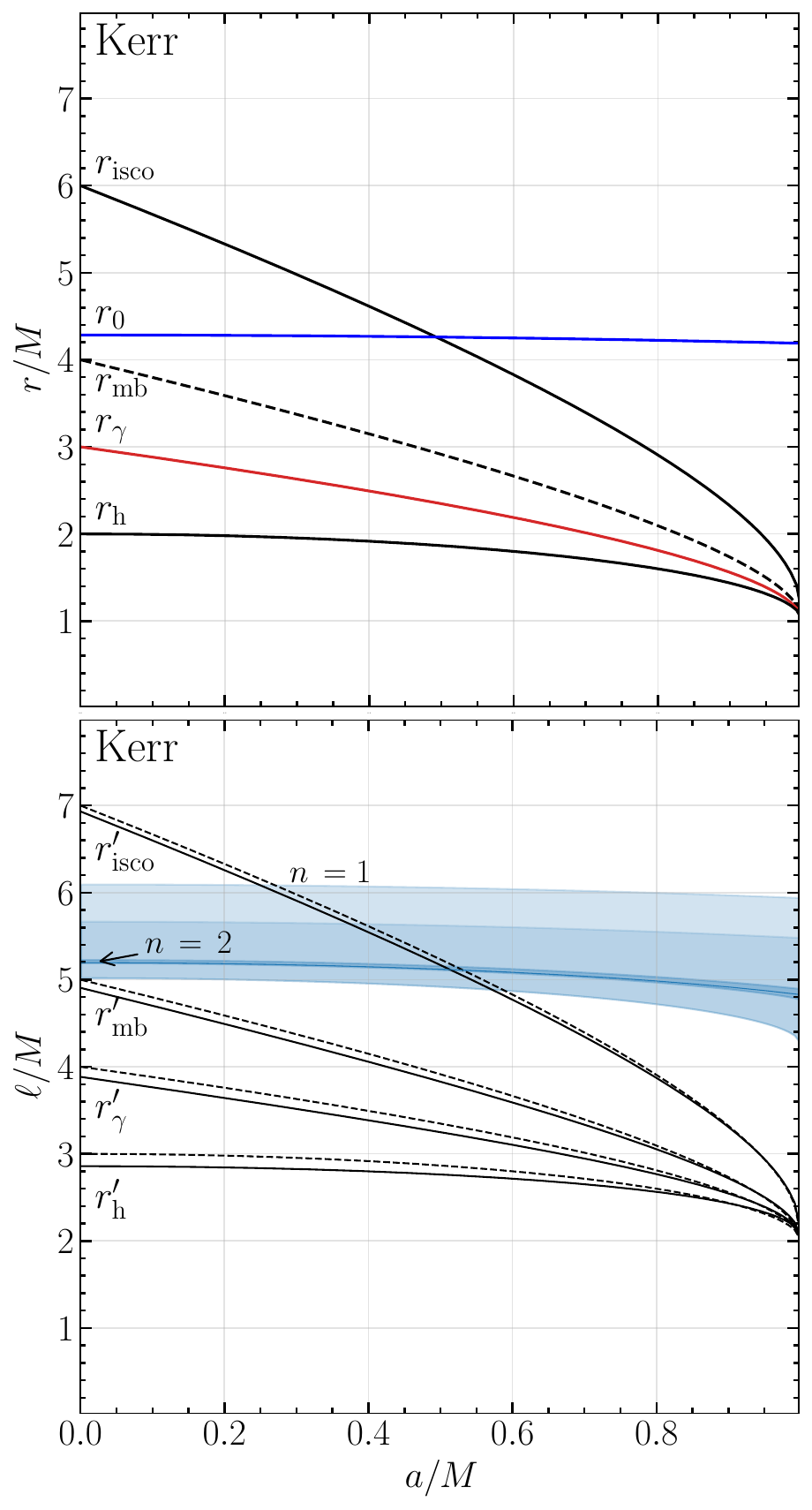}
        \caption{Similar as Fig. \ref{fig:structure_BH} and Fig. \ref{fig:structure_image}, but for the the Kerr spacetime, parametrized with a spin parameter $a/M$, with the accretion disk viewed from the BH spin axis, aligned with the accretion disk spin axis. Characteristic radii represent the equatorial slice of the Kerr spacetime \citep{Bardeen1972}.
   }
    \label{fig:structure_kerr}
\end{figure}

Additionally, we plot a line $r_0$, representing the effective radius of emission $r_{\rm e}$, that would result in a direct image of ring with a radius of $ \ell_0 = 3\sqrt{3} M$, which is the critical curve radius in Schwarzshild spacetime. The charge parameter $q$ has a very limited influence on the value of $r_0$, which remains in the $4.3 \pm 0.1 M$ range for all the considered spacetimes. This is in contrast to some other characteristic radii, for example between $q=0$ and $q=1$ ISCO location in RN spacetime changes by over 30\%, while $r_0$ changes by less than 4\%. Moreover, Kerr spacetime viewed face-on also gives consistent $r_0$ for all considered values of spin ($0 \le a/M < 1)$, Fig. \ref{fig:structure_kerr}. This shows that in a hypothetical case of a direct emission ring measurement perfectly consistent with the Schwarzschild critical curve, one \textit{can not} claim a metric test (constraint on $a$ or $q$) without additional strong assumptions on the accretion model (what $r_{\rm e}$ is allowed for which spacetimes). On the other hand, one could claim a constraint on the object compactness, broadly independent on the underlying spacetime metric. In case of the EHT observations of M\,87$^*$, assuming mass and distance priors \citep{EHT2019p6}, one would find the M\,87$^*$ upper bound on compactness limited by $r_{\rm e}/M = 4.5 \pm 0.7$, where uncertainties are dominated by the object mass and distance uncertainties \citep{Gebhardt2011,EHT2019p6}.

\begin{figure*}[t!]
    \centering

     \includegraphics[width=0.325\linewidth]{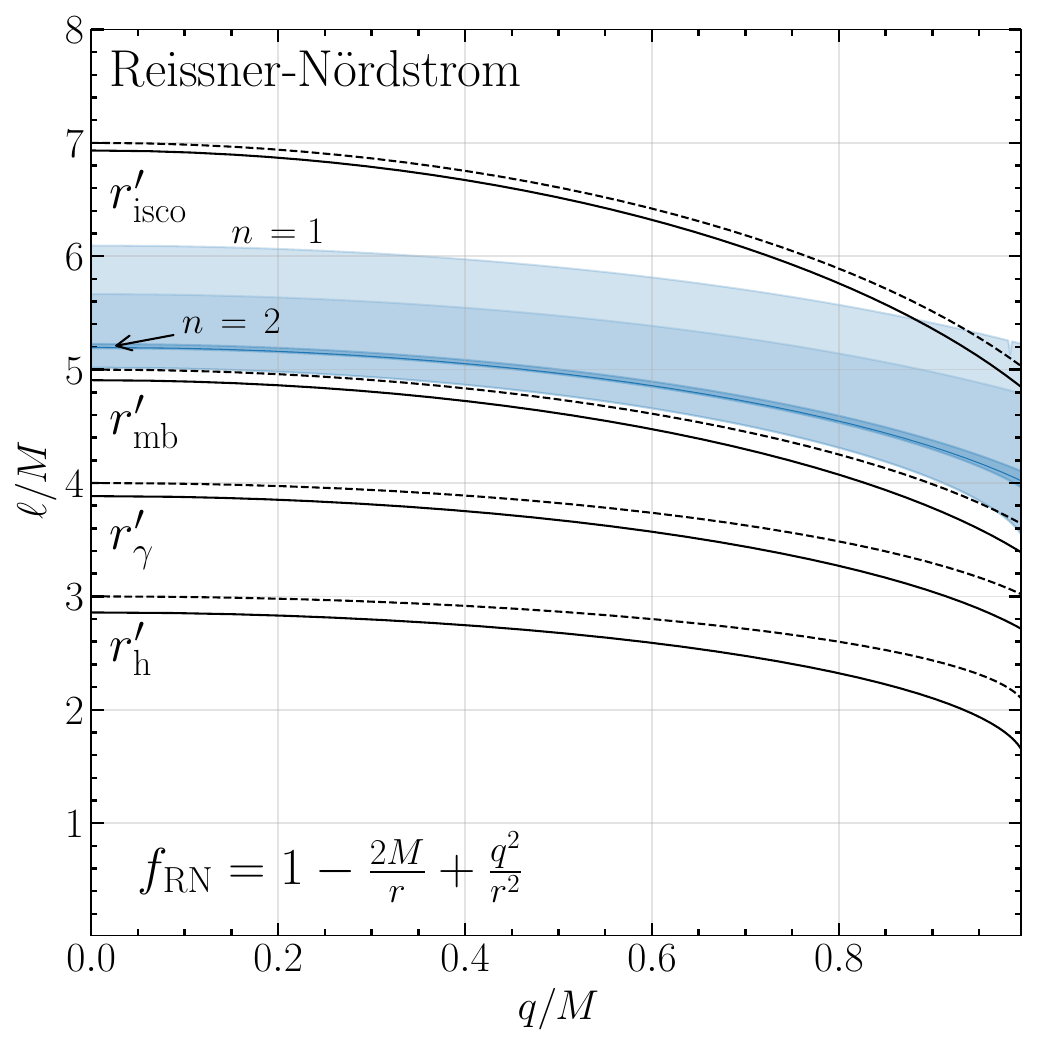}
     \includegraphics[width=0.325\linewidth]{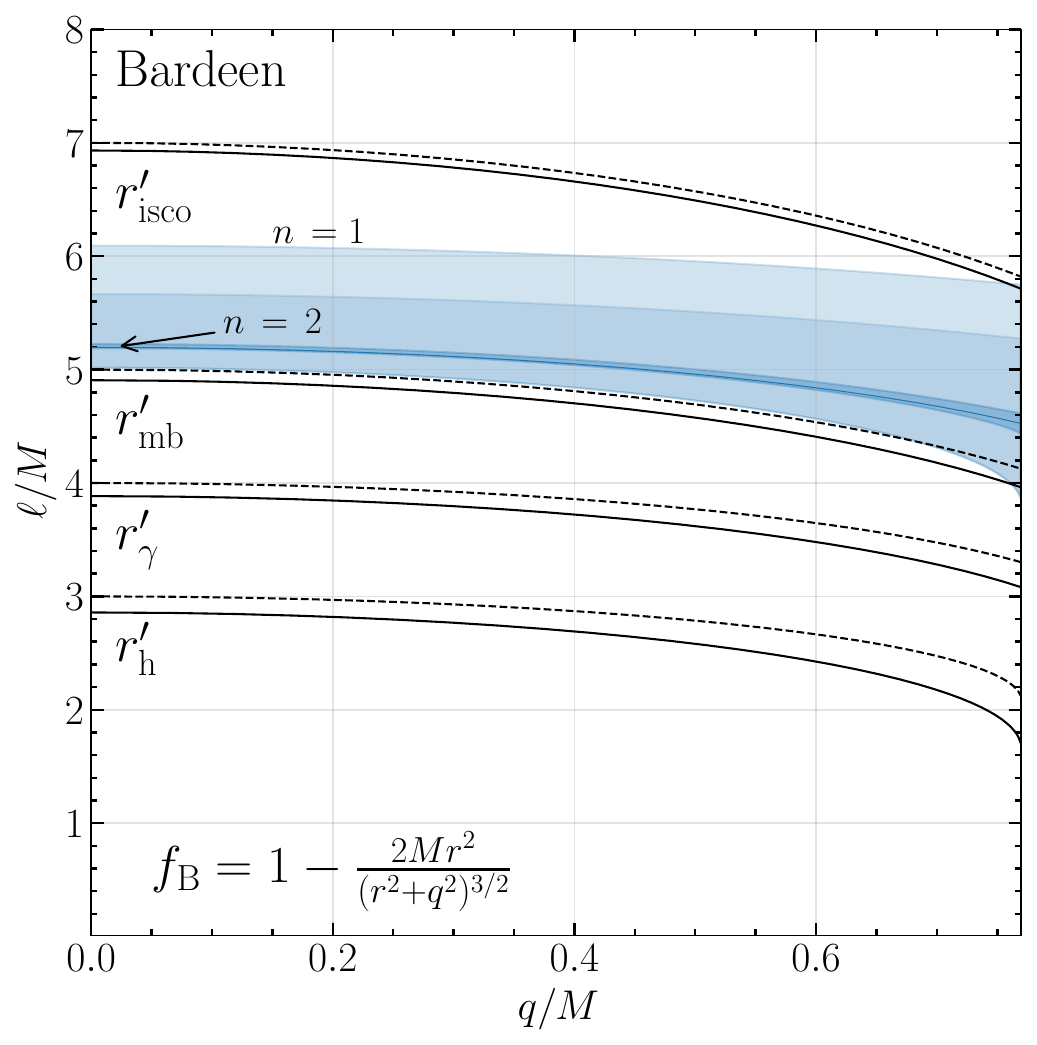}
     \includegraphics[width=0.325\linewidth]{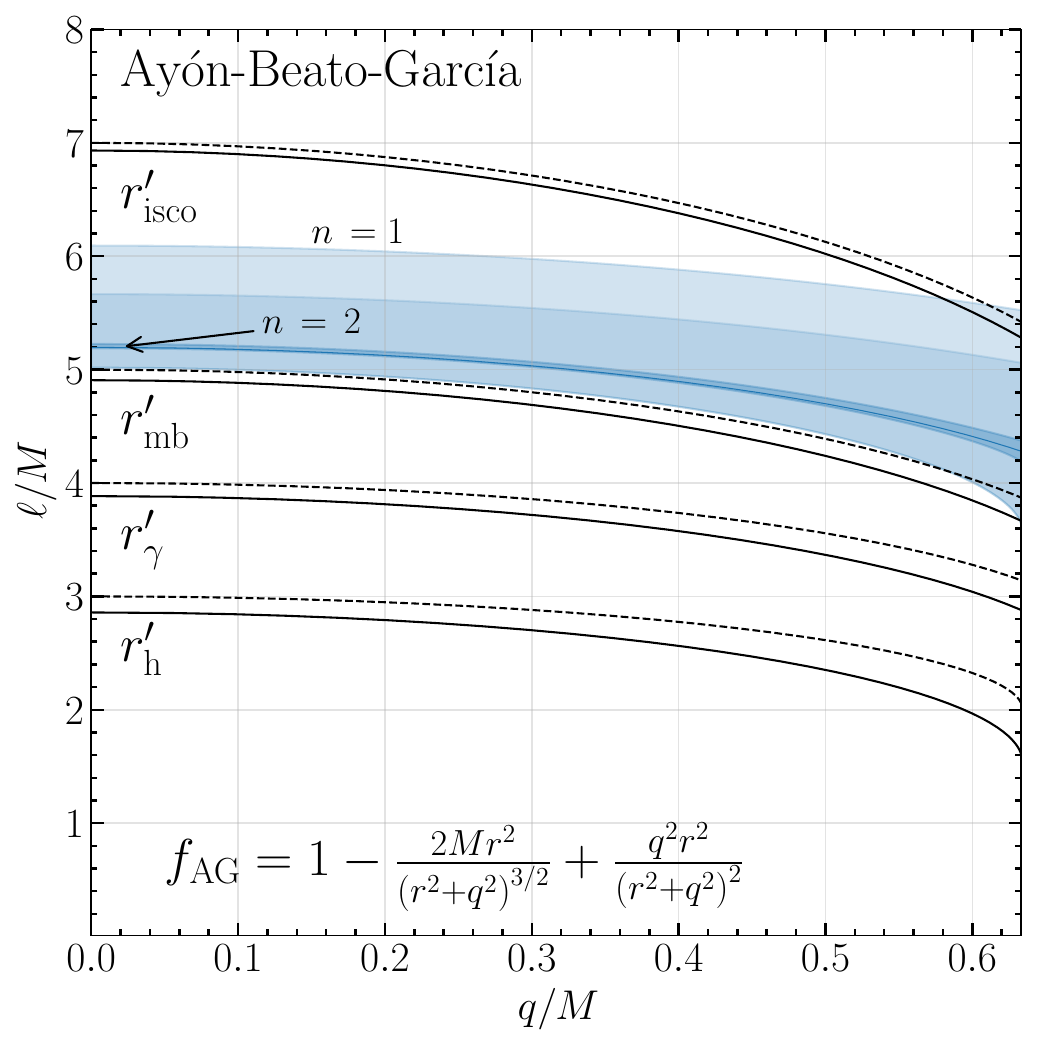}
     \includegraphics[width=0.325\linewidth]{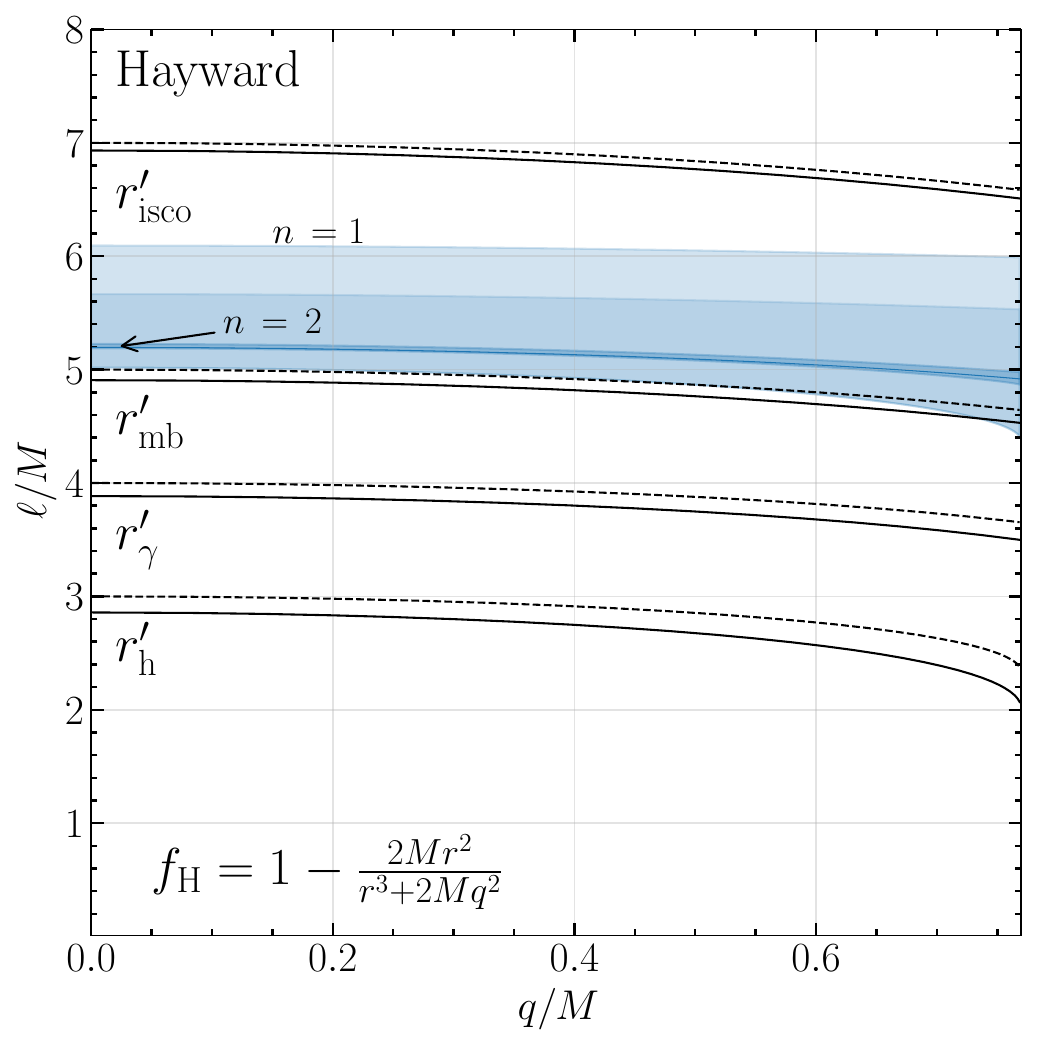}
     \includegraphics[width=0.325\linewidth]{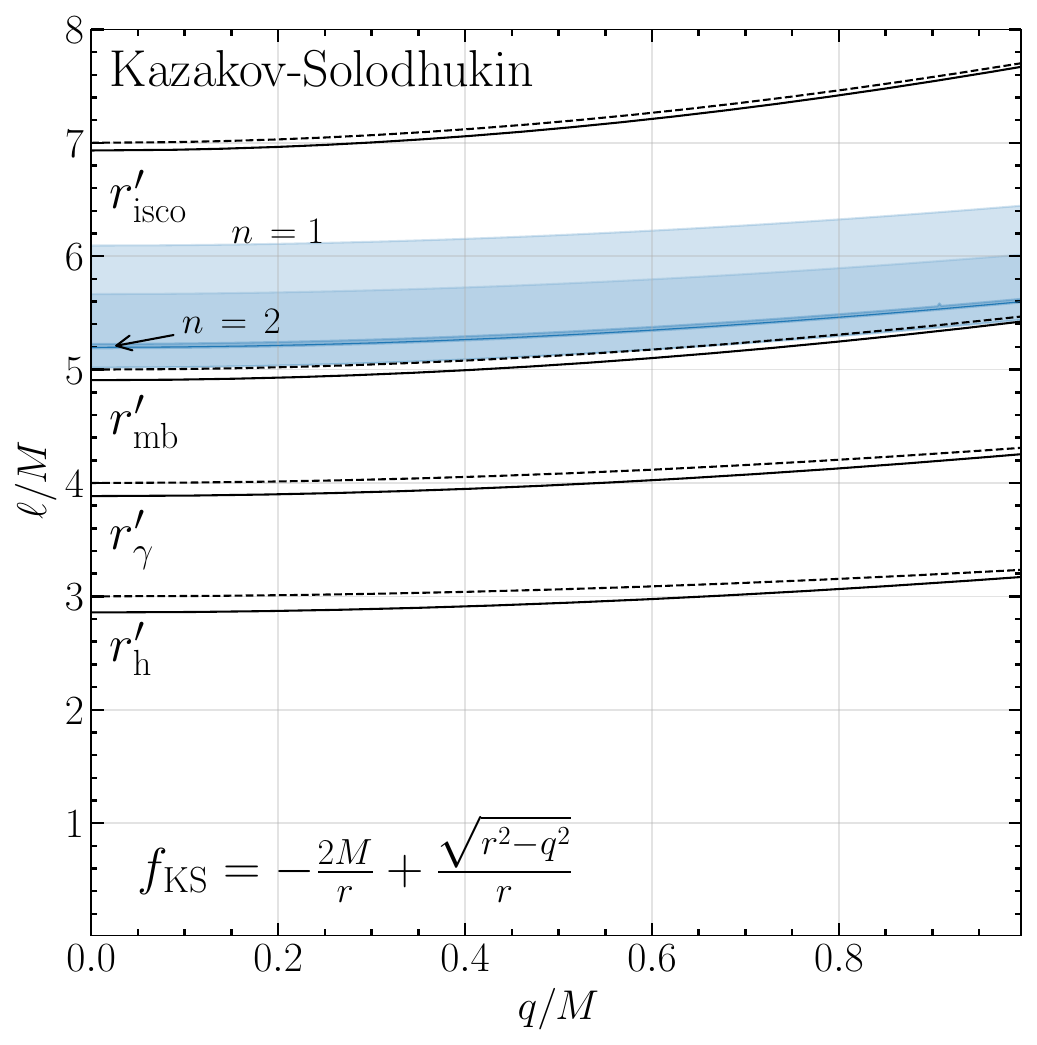}
     \includegraphics[width=0.325\linewidth]{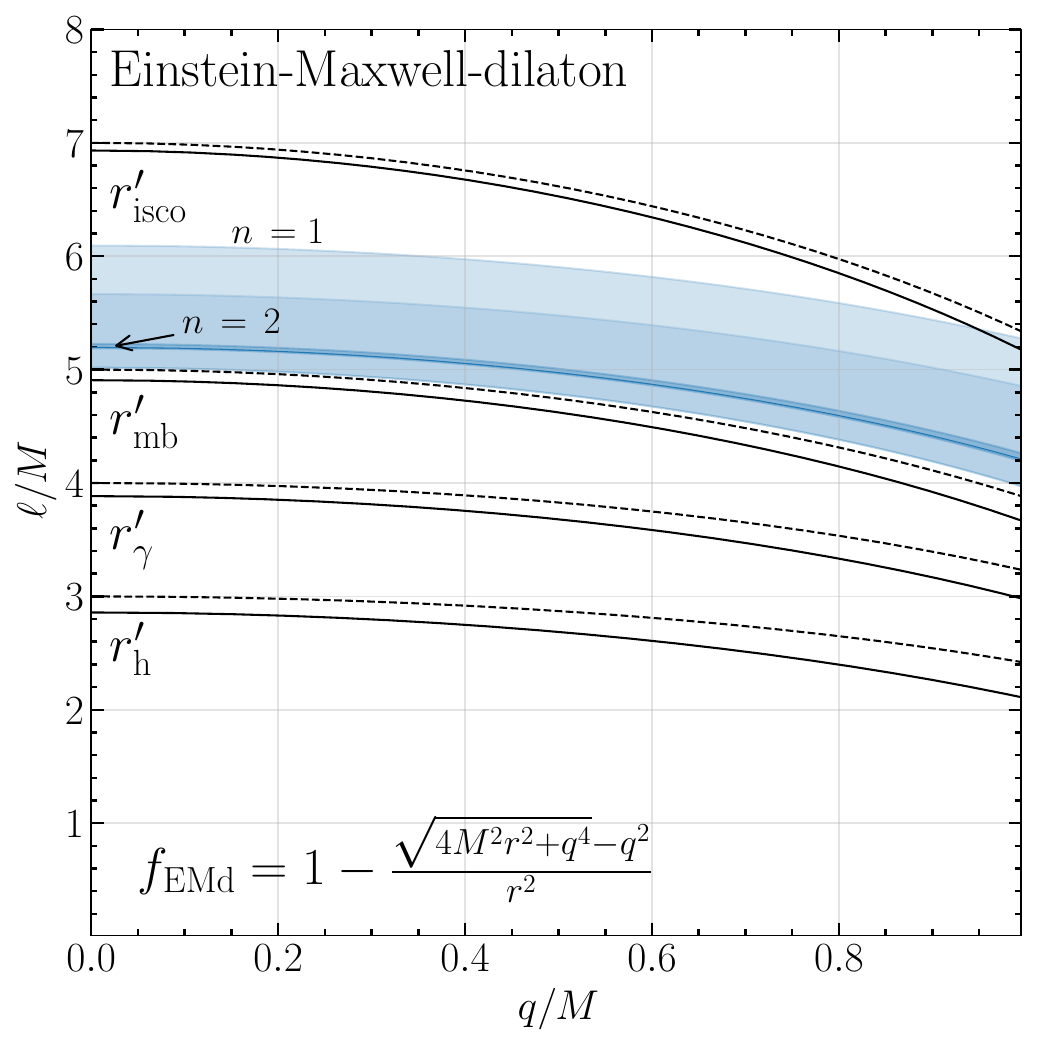}
    \caption{Image domain structure of the selected spacetimes. Locations of characteristic radii are denoted with prime. Dashed lines represent the $\ell = r_{\rm e} + M$ formula. Blue bands indicate the range of possible radii for the observed photon rings. For $n=1$ lighter band represents full range of $r_{\rm h} < r_{\rm e} < \infty$, and darker band assumes $r_{\rm e} < 10 M$. The narrow $n=2$ band covers whole range of $r_{\rm e}$ and contains the critical curve.
   }
    \label{fig:structure_image}
\end{figure*}

\section{Photon Ring Calculations}
\label{sec:calculations}

For the case of Kerr spacetime, the null geodesic structure has been studied extensively, and geometry of the photon rings can be calculated analytically without resorting to ray-tracing or explicit numerical integration of the geodesic equation, see, e.g., \citep{Gralla2019,Gralla_lensing}. In a spherically symmetric case of a metric given by Eq. \ref{eq:metric} with two unspecified functions $f(r)$ and $g(r)$ we resort to the numerical integration of the deflection angle $\delta$ to find the location of the photon rings, noticing that the angular path between the emission point and the  distant observer is equal to
\begin{equation}
    \delta_n = \frac{\pi}{2} + n \pi \, 
\end{equation}
in our simplified model of the system geometry (see Fig. \ref{fig:introfigure}).
Specific angular momentum $\ell = -u_\phi/u_t$, is conserved along geodesics due to the Killing symmetries of the static axisymmetric spacetimes, and corresponds to the impact parameter of the photon as seen on a screen of a distant observer, Fig. \ref{fig:introfigure}. Hence, using the deflection angle formula, we can write down an implicit relation between the effective emission radius $r_{\rm e}$ and the impact parameter of the $n$-th photon ring $\ell_n$ for a given metric $g_{\mu \nu}$
\begin{align}
  F_n( \ell_n ; r_{\rm e}, g_{\mu \nu}) &= \Bigg| \int\displaylimits^{r_{\rm e}}_\infty \frac{u^{\phi}}{u^r} \dd r \Bigg|   - \delta_n \nonumber \\
  &= \Bigg| \int\displaylimits^{r_{\rm e}}_\infty 
  \frac{\dd r}{\sqrt{ \frac{r^2}{g(r)} \left[ \frac{r^2}{\ell_n^2} - f(r) \right]}} \Bigg|   -\delta_n \ .
  \label{eq:find_photonring}
\end{align}
The $n$-th photon ring corresponds to $F_n(\ell_n; r_{\rm e}, g_{\mu \nu}) = 0$ if between the emission and the distant observer trajectory does not go through a pericenter of the null geodesic. In other case, the pericenter 
radius $r_{\rm p}$ can be found from the radial four-velocity $u^r = 0$ condition, the total angular path along the fully extended photon trajectory is then
\begin{equation}
    \Delta \phi_{\rm tot} = 2 \Bigg| \int\displaylimits^{r_{\rm p}}_\infty \frac{u^{\phi}}{u^r} \dd r \Bigg|
\end{equation}
and the impact parameter of the photon ring $\ell_n$ can be found by comparing $F'_n$ to zero
\begin{equation}
    F'_n( \ell_n ; r_{\rm e}, g_{\mu \nu})  = \Delta \phi_{\rm tot} - F_n( \ell_n ; r_{\rm e}, g_{\mu \nu}) - 2 \delta_n \ .
    \label{eq:find_photonring_pericenter}
\end{equation}
That is the case for $n\!=\!1,2$ (but not for $n\!=\!0$) trajectories shown in Fig. \ref{fig:introfigure}. In the figure we indicate the locations of the pericenters $r_{\rm p}$. A practical criterion for using Eq. \ref{eq:find_photonring_pericenter} rather than Eq. \ref{eq:find_photonring} is that $r_{\rm p}$ exists and $\Delta \phi_{\rm tot} < 2 \delta_n$, so that the pericenter is located between the emitter and the observer.
The critical curve location can be found without resorting to numerical integration as
\begin{equation}
    \ell_{\rm c} = \frac{r_\gamma}{\sqrt{f(r_\gamma)}} \ ,
\end{equation}
where the photon sphere radius $r_\gamma$ is given as a solution to Eq. \ref{eq:rgamma}.
We illustrate the image domain characteristic radii for each spacetime in Fig. \ref{fig:structure_image}. Here, the location of direct images of characteristic radii from Fig. \ref{fig:structure_BH} are shown (black continuous lines) and denoted with a prime. In particular, $r_{\rm h}'$ can be associated with the \textit{central brightness depression} of \citep{Bronzwaer2021} or with the \textit{inner shadow} of \citep{Chael2021}. Dashed lines represent the approximated location of direct images calculated with the Eq. \ref{eq:adding_one}. This trivial formula is reasonably accurate for all considered spacetimes, yielding at most $\sim 20\%$ deviation in case of $r_{\rm h}'$ (that is $r_{\rm e} \gtrapprox r_{\rm h}$) and a large charge, and much less in all other cases. This once again shows that the direct image is not a robust and direct probe of the metric, and instead it rather unambiguously probes the location of the emission $r_{\rm e}/M$ independently of the details of the metric (note that the accuracy of this measurement is ultimately limited by the uncertainties in the distance $D$ and simplified geometry of the model). The figure also indicates possible range of radii of the $n=1$ and $n=2$ photon rings as seen by a distant observer. Similar plot for the Kerr spacetime parametrized with spin $a/M$ is given in the bottom panel of the Fig. \ref{fig:structure_kerr}. We only consider the case of the Kerr BH spin aligned along the the line of sight, and for the calculations we follow procedures outlined in \citep{Gralla_lensing}. It is now evident that while the direct image approximately reproduces the structure of the emission (shifted by $1\,M$ following Eq. \ref{eq:adding_one}), photon rings transfer different emission radii into small range of radii in the observer's plane, see also transfer functions example discussed in Section \ref{sec:astrophysical_constraints}. For all considered cases $n=1$ approximates the critical curve within 15\%, while $n=2$ closer than 2.2\%, independently of the emission profile. Hence, photon rings are sensitive to the spacetime geometry.

\begin{figure}[t!]
    \centering

     \includegraphics[width=0.95\linewidth]{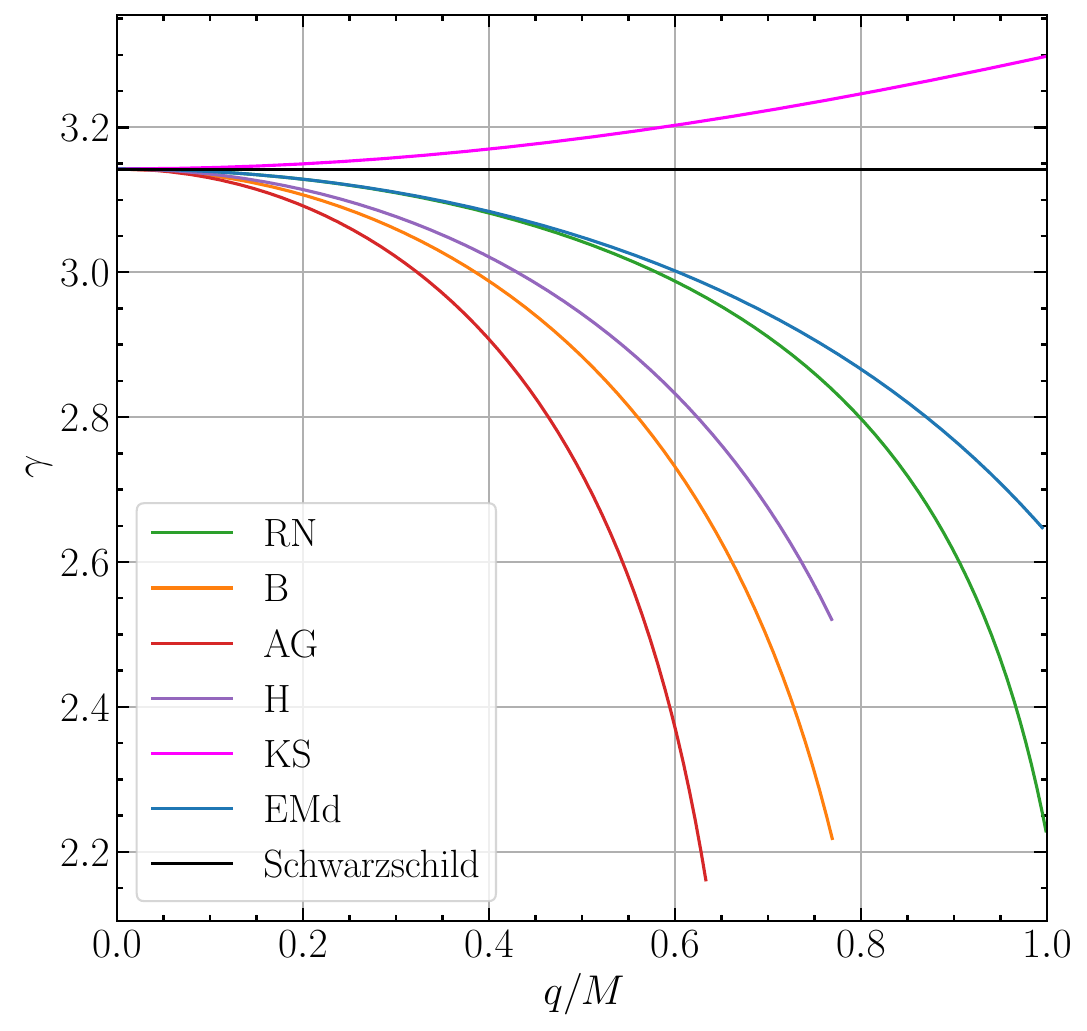}
        \caption{Lyapunov exponents $\gamma$ calculated for the considered families of spacetimes with charges $q$. For Schwarzschild case we find $\gamma = \pi$. For spacetimes with smaller $\gamma$, we expect relatively more flux in the subsequent photon rings. For $\gamma = 2.45$ the flux ratio difference with respect to the Schwarzschild case corresponds to a factor of 2.
   }
    \label{fig:LyapunovG}
\end{figure}

\begin{figure*}[t!]
    \centering
\includegraphics[width=0.999\linewidth]{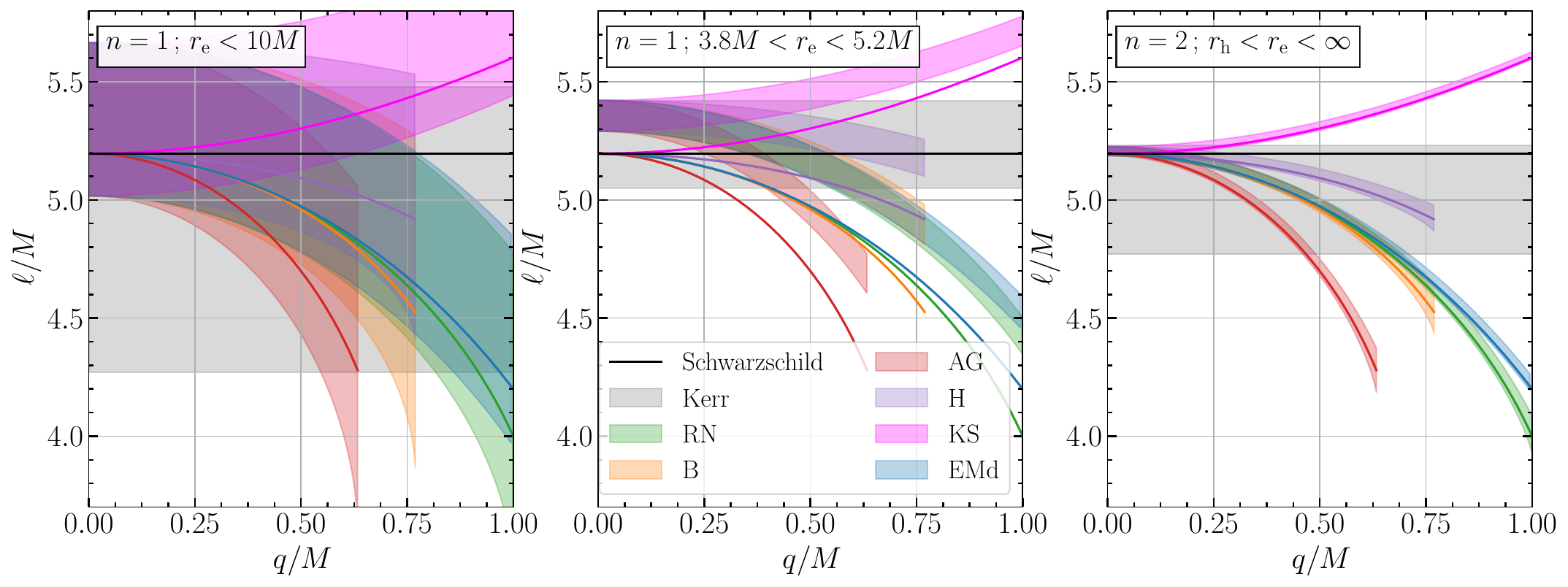}
        \caption{\textit{Left:} range of possible photon ring radii in the observer's plane, corresponding to $n=1$ photon ring, assuming the emission radius $r_{\rm e}< 10 M$. Solid lines correspond to critical curves in different spacetimes. Gray band represents $n=1$ photon ring radii for Kerr metric with spin $0 \le a/M < 1$.
        \textit{Middle:} same as left, but assuming $3.8M < r_{\rm e}< 5.2 M$ following the EHT constraints on the M\,87$^*$. \textit{Right:} same, but for the $n=2$ photon ring and $r_{\rm h} < r_{\rm e}< \infty$.
   }
    \label{fig:1and2_rings}
\end{figure*}

Another interesting characteristic of a spacetime, related to its photon ring structure, is provided by the Lyapunov exponents $\gamma$ \citep{Johnson2020}, quantifying the instability of the bound photon shell geodesics. In case of spherically symmetric spacetimes, if photon's radial position is perturbed from the spherical geodesic $r_\gamma$ radially by an infinitesimal $\delta r_0$, after $n$ half-orbits the photon will appear at the radius $\delta r_n$ away from $r_\gamma$ and
\begin{equation}
    \delta r_n = e^{\gamma n} \delta r_0 .
\end{equation}
We evaluate the Lyapunov exponents numerically as $\gamma \approx \ln (\delta r_1/\delta r_0)$ for the considered families of spacetimes, see Fig. \ref{fig:LyapunovG}. For Schwarzschild case $\gamma= \pi$ \citep{Luminet1979}. Lyapunov exponents control the relative flux in subsequent photon rings, and hence constitute an observable quantity. While for Schwarzschild case flux in subsequent rings is extinguished by a factor of $e^{-\pi} \approx 0.04$, for a spacetime with $\gamma = 2.45$ we obtain a twice less severe extinction -- in such a spacetime the photon rings are expected to be significantly brighter.

\section{Photon Ring Metric Tests}

Is the photon ring insensitivity to the accretion model sufficient to perform a robust test of the metric? In Fig. \ref{fig:1and2_rings} we show the possible range of photon ring radii for all considered spacetimes. For the first photon ring $n=1$ (left panel) all bands corresponding to charged BH spacetimes significantly overlap with the full possible range of the first photon ring radius in Kerr spacetimes. Hence, only some limited combinations of large charge and particularly small (or large, for the KS spacetime) emission radius would allow us to conclude inconsistency with the Kerr hypothesis. It makes a big difference if we can further restrict the emission radius $r_{\rm e}$. As an example, the middle panel of Fig. \ref{fig:1and2_rings} shows the $n=1$ photon ring location assuming that $r_{\rm e} = 4.5 \pm 0.7 M$, which is consistent with the constraints on the emission radius based on the EHT observations of M\,87$^*$ reported in Section \ref{sec:spacetimes}. Although the critical curve is no longer within the range of possible $n=1$ radii, it is apparent that the measurement of a first photon ring combined with a prior on $r_{\rm e}$ from the measurement of the direct image radius could lead to constraints on charges as well as to the Kerr hypothesis test. The situation appears even more encouraging in the case of the $n=2$ photon ring, right panel of Fig. \ref{fig:1and2_rings}.

\begin{figure}[t!]
    \centering

     \includegraphics[width=0.98\linewidth]{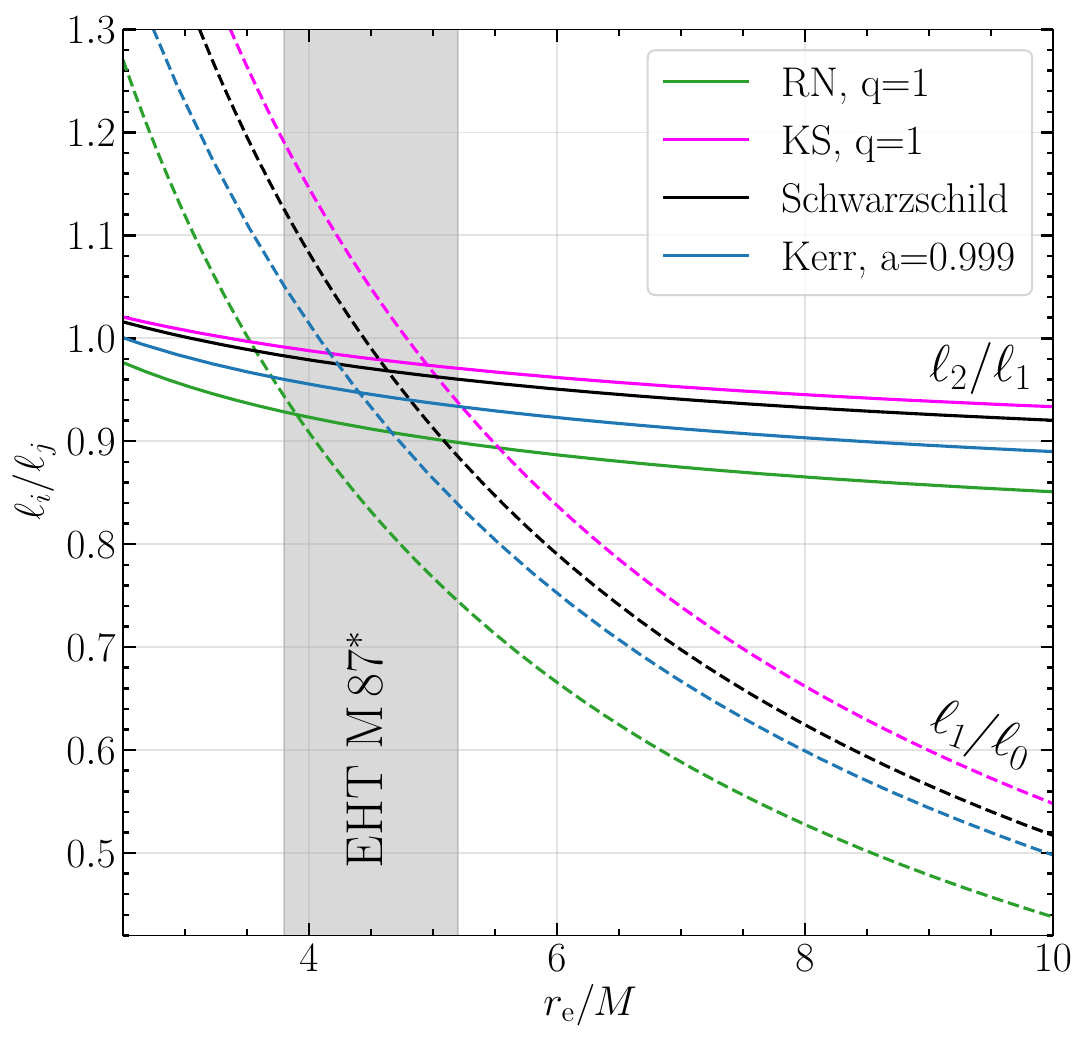}
        \caption{Ratios of photon ring radii for selected spacetimes as a function of the emission radius $r_{\rm e}$. The shaded region corresponds to the $r_{\rm e}$ limits inferred from the EHT results \citep{EHT2019p1}.
   }
    \label{fig:ratios_re}
\end{figure}

So far we neglected the impact of the object mass and distance uncertainties on the photon ring radius measurement. However, the related uncertainties may be significant. In case of the M\,87$^*$ assumed priors resulted in $M/D$ uncertain to about 12\% \citep{Gebhardt2011,EHT2019p6}. Clearly, this level of uncertainty could already dominate the error budget and disable metric tests with photon rings, as it is comparable with the whole range of possible photon ring radii for all considered metrics shown in Fig. \ref{fig:1and2_rings}. It has been suggested by \citep{Broderick2021} that the Kerr metric spin could be probed using a dimensionless ratio of the photon rings radii, independently of the $M/D$ scaling factor. Similarly, we could envision using such ratios to put constraints on the non-Kerr spacetime models. In Fig. \ref{fig:ratios_re} we show the ratios of radii of second to first ($\ell_2/\ell_1$) and first to zeroth (direct image; $\ell_1/\ell_0$) photon rings as a function of the emission radius $r_{\rm e}$. For clarity, we only compare charged BH spacetimes with largest deviation from the Schwarzschild spacetime, that is RN and KS with $q=1$, as well as the Kerr spacetime with a large spin ($a/M = 0.999$). The remaining problem is that the $r_{\rm e}$ parameter is not known. However, as we discussed in Section \ref{sec:spacetimes}, the size of the direct image $n=0$ can give us a very decent, metric independent, estimation of the $r_{\rm e}$, within the $M/D$ uncertainties. Hence, a combined method, using $M/D$ independent $\ell_2/\ell_1$ ratio and $r_{\rm e}$ prior (that is not free of the $M/D$ uncertainties), could be employed. Given the uncertainties in case of the M\,87$^*$ observations we assume $r_{\rm e} = 4.5 \pm 0.7 M$, shaded region in Fig. \ref{fig:ratios_re}. Then it is clearly possible to constrain charges and test Kerr hypothesis, particularly with the $\ell_2/\ell_1$ ratio. It may even be possible to constrain the spin value, following \citep{Broderick2021}.

\section{Conclusions}

We studied structure of lensed BH images in several spherically symmetric BH spacetimes representing deviation from the Schwarzschild spacetime parametrized with a charge $q$, as well as in the Kerr spacetime viewed along the spin axis. The radius of the direct image of the accretion disk is not directly sensitive to the spacetime metric. Instead, it is sensitive to the effective radius of the emission, which in turn may be be sensitive to the metric, but it is also strongly dependent on the astrophysics of the accretion flow. Based on the M\,87$^*$  direct image radius reported by the EHT \citep{EHT2019p1}, we measure the  M\,87$^*$ effective emission radius to be $4.5 \pm 0.7 M$, independently of the underlying metric, indicating extreme compactness of this object. Higher order photon rings are sensitive to the spacetime metric, and may be used in the future to probe the metric independently (or with a strongly bounded dependence) of the underlying model of the accretion flow. We have demonstrated how first or second photon ring measurement could be used to constrain BH charges in different spacetime models, particularly when such measurement is coupled with a prior on $r_{\rm e}$ from the direct image radius measurement.

Presented results assume a very simplified setup, ignoring the spin of charged BHs, assuming face-on view of the accretion disk, and emission from an equatorial thin ring. In order to properly interpret observational measurements, a more general model needs to be employed. At the moment, theory is a number of years ahead of the observations, so these much more detailed and technical calculations may be performed in the future work -- the relevant conclusion is that the photon ring structure is only weakly influenced by the assumed source model. While accounting for the inclination increases the number of degrees of freedom in the model, it also unlocks the possibility of performing metric tests based on non-circular photon ring geometry \citep{Gralla_n2,Paugnat2021}.

Finally, we sensitise the readers against the straightforward interpretation of the EHT measurements of the direct image properties (diameter, circularity) as statements about the critical curve geometry. These are different entities, and in order to make a connection between them, a model of the emission source structure needs to be employed. It should be clearly stated when the inferred properties are depending on such a model, and what are its restrictions. It is both interesting and useful to calibrate EHT direct image observations with a limited set of theoretical models of the accretion flow, such as GRMHD simulations in \citep{EHT2019p5}, while recognizing the limitations of this framework. Assumptions should be explicitly stated when a measurement of any physical parameter is performed. 

\begin{acknowledgments}
 We thank Zack Gelles, Sam Gralla, Michael Johnson, Prashant Kocherlakota, Roman Konoplya, Alex Lupsasca, Daniel Palumbo, Bart Ripperda, and last but not least Frederic Vincent for enlightening discussions and comments. We also thank Alexandra Elbakyan for her contributions to the open science. This work was supported by the Black Hole Initiative at Harvard University, which is funded by grants from the John Templeton Foundation and the Gordon and Betty Moore Foundation to Harvard University.
\end{acknowledgments}


\bibliography{transfer}

\end{document}